\documentclass[graybox]{svmult}

\usepackage{mathptmx}       
\usepackage{helvet}         
\usepackage{courier}        
\usepackage{type1cm}        
\usepackage[psamsfonts]{amssymb}
\usepackage{amsmath}
\usepackage{amsfonts}
\usepackage{bm}

\usepackage{makeidx}         
\usepackage[dvipdfm]{graphicx}        
\usepackage{multicol}        
\usepackage[bottom]{footmisc}
\usepackage{bm}

\makeindex             

\begin{document}

\title*{Characterizing financial crisis by means of the three states random field Ising model}
\titlerunning{Characterizing financial crisis by means of the three states RFIM}
\author{Mitsuaki Murota and Jun-ichi Inoue}
\institute{Mitsuaki Murota \at Graduate School of Information Science and Technology, 
Hokkaido University, N14-W-9, Kita-ku, Sapporo 060-0814, Japan \email{murota@complex.ist.hokudai.ac.jp}
\and Jun-ichi Inoue \at Graduate School of Information Science and Technology,
Hokkaido University, N14-W-9, Kita-ku, Sapporo 060-0814, Japan \email{jinoue@cb4.so-net.ne.jp}}
%
%
\maketitle

\abstract{
We propose a formula of time-series prediction 
by means of three states random field Ising model (RFIM). 
At the economic crisis due to disasters or international disputes, 
the stock price suddenly drops. The macroscopic phenomena 
should be explained from the corresponding microscopic view point  because 
there are existing a huge number of active traders behind the crushes. 
Hence, here we attempt to model the artificial financial market 
in which each trader $i$ can choose his/her decision 
among `buying', `selling' or `staying (taking a wait-and-see attitude)', 
each of which corresponds to a realization of the three state Ising spin, namely,  
$S_{i}=+1$, $-1$ and $S_{i}=0$, respectively. 
The decision making of traders is given by 
the Gibbs-Boltzmann distribution 
with the energy function. 
The energy function contains three distinct terms, 
namely, the ferromagnetic two-body interaction term (endogenous information), 
random field term as external information (exogenous news), 
and chemical potential term which controls the number of traders who 
are watching the market calmly at the instance. 
We specify the details of the model system from the past financial market data 
to determine the conjugate hyper-parameters and draw each parameter flow as a function of time-step. 
Especially we will examine to what extent 
one can characterize the crisis 
by means of a brand-new order parameter --- `turnover' --- 
which is defined as the number of active traders who post their decisions  
$S_{i}=\pm 1$, instead of $S_{i}=0$. 
}

\section{Introduction}
\label{sec:Intro}
Individual human behaviour including human mental state is an attractive topic 
for both scientists and engineers. 
However, it is still extremely difficult for us to tackle 
the problem by making use of scientifically reliable investigation. 
This is because there exists quite large person-to-person fluctuation 
in the observation of individual behaviour.  

On the other hand, in our human `collective' behaviour instead of individual, 
we sometimes observe several universal facts 
which seem to be suitable materials for computer scientist 
to figure out the phenomena through sophisticated approaches such as agent-based simulations. 
In fact,  collective behaviour of interacting agents such as flying birds, moving insects or 
swimming fishes shows highly non-trivial properties. 
As well-known especially in the research field of engineering, 
 as a simplest and effective algorithm in computer simulations 
 for flocks of intelligent agents, say,  animals such as starlings, 
 the so-called BOIDS founded by Reynolds \cite{Reynolds}
 has been widely used not only in the field of computer graphics but also in 
 various other research fields  including ethology, physics, 
 control theory, economics, and so on \cite{Makiguchi}. 
 The BOIDS simulates the collective behaviour of animal flocks 
 by taking into account only a few simple rules for each 
 interacting `intelligent' agent. 

In the literature of behavioral economics \cite{Kahneman}, a concept of the so-called information cascade is well-known as a result of 
such human collective behaviour. 
This concept means that at the financial crisis, 
traders tend to behave according to the `mood' (atmosphere) in society (financial market) and 
they incline to take rather `irrational' strategies in some sense. 

Apparently, one of the key measurements  
to understand the information cascade 
is `correlation' between ingredients 
in the societies (systems). 
For instance, in particular for financial markets, cross-correlations between stocks, 
traders are quite important to figure out the human collective phenomena. 
As the correlation could be found in various scale-lengths, 
from macroscopic stock price level to microscopic trader's level,   
the information cascade also might be 
observed `hierarchically' in such various scales 
from prices of several stocks to  
ways (strategies) of trader's decision making. 

Turning now to the situation of Japan, 
after the earthquake on 11th March 2011, Japanese NIKKEI stock market quickly responded to the crisis and quite a lot of traders 
sold their stocks of companies whose branches or plants are located in that disaster stricken area. 
As the result, the Nikkei stock average suddenly drops after the crisis \cite{ISI,IHSI}.

It might be quite important for us to make an attempt to bring out more `microscopic' useful information, 
which is never obtained from the averaged macroscopic quantities such as stock average, 
about the market.  As a candidate of such `microscopic information', we can use the (linear) correlation coefficient 
based on the two-body interactions between stocks \cite{Mantegna,Anirban}. 
To make out the mechanism of financial crisis, it might be helpful for us to visualize 
such correlations in stocks and compare the dynamical behaviour of the correlation before and after crisis. 

In order to show and explain 
the cascade, we visualized the correlation of each stock in two-dimension \cite{ISI,IHSI}. 
We specified each location of $N$ stocks from a given set of the 
$N(N-1)/2$ distances by making use of the so-called multi-dimensional scaling (MDS) \cite{MDS}. 
\begin{figure}[ht]
\begin{center}
\includegraphics[width=11cm]{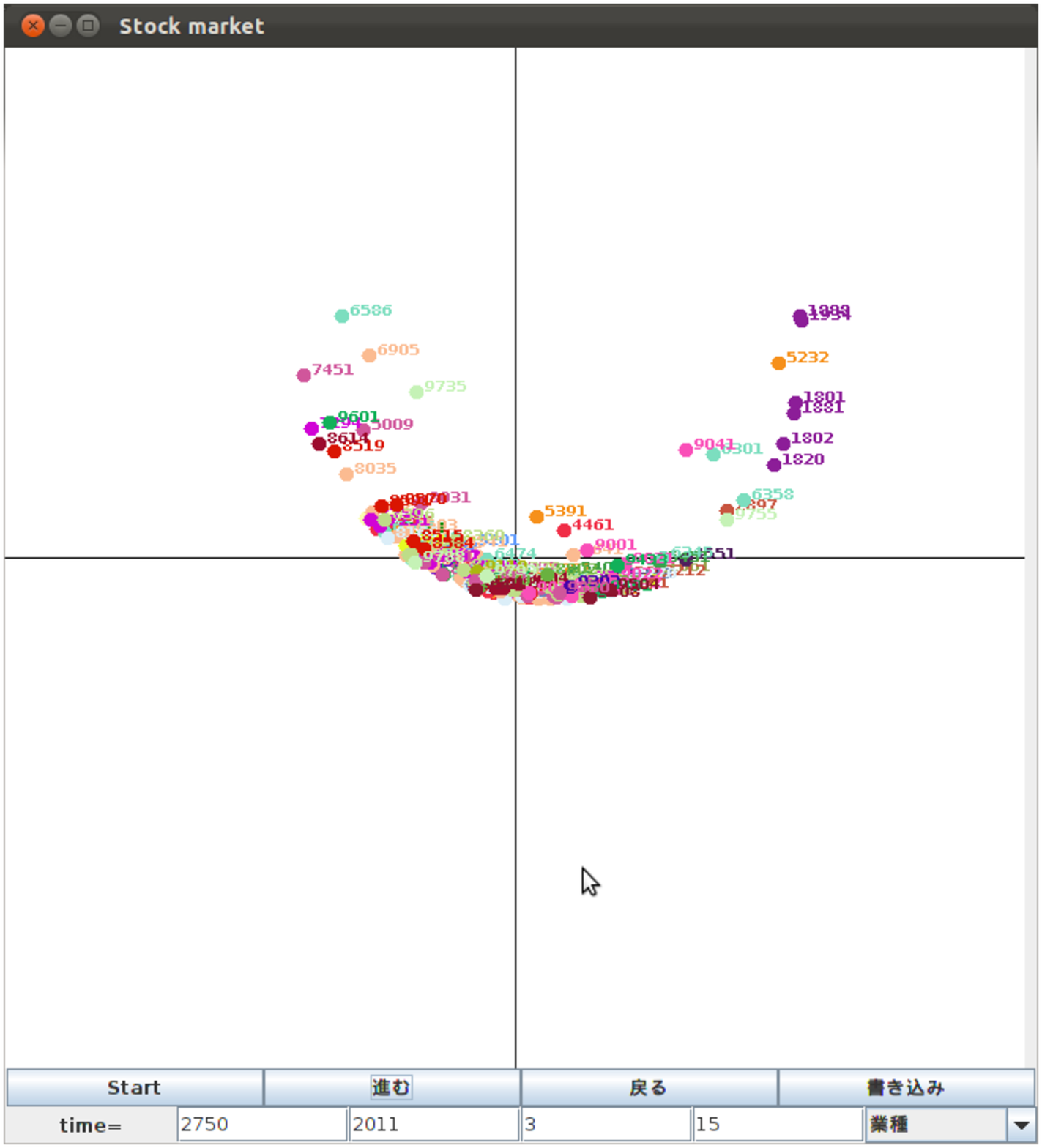}
\end{center}
\caption{\footnotesize 
Two-dimensional plot by means of the MDS. 
We picked up 200-stocks including 
the so-called TOPIX Core30 and 
the Nikkei stock average as 
empirical data set (the data set was taken from Yahoo! finance \cite{Yahoo}). 
The figure is just after the earthquake (15th March 2011) (see \cite{ISI,IHSI} 
for the details). 
The curious shape of cluster appears after the crisis. 
}
\label{fig:fg_muro1}
\end{figure} 
\mbox{}

On the other hand, 
the macroscopic phenomena 
should be explained from the corresponding microscopic view point  because 
there are existing a huge number of active traders behind the crushes. 
In the reference \cite{IHSI}, 
we proposed a theoretical framework to predict several time-series 
simultaneously by using cross-correlations in financial markets. 
The justification of this assumption was numerically checked for the empirical Japanese stock data, 
for instance, those around 11 March 2011, 
and for foreign currency exchange rates around Greek crisis in spring 2010. 

However, in the previous study \cite{IHSI}, inspired by the study of Kaizoji \cite{Kaizoji2000}, 
we utilized Ising model and assumed that each trader does not stay at all for trading. 
Apparently, it is not realistic situation for trader's decision making. 
Hence, here we attempt to model the artificial financial market 
in which each trader $i$ can choose his/her decision 
among `buying', `selling' or `staying (taking a wait-and-see attitude)', 
each of which corresponds to a realization of the three states Ising spin, namely,  
$S_{i}=+1$, $-1$ and $S_{i}=0$, respectively. 
Especially we will examine to what extent 
one can characterize the crisis 
by means of an order parameter --- `turnover' --- which is defined as the number of `active traders' who post their decision 
$S_{i}=\pm 1$, instead of $S_{i}=0$. 

This paper is organized as follows. 
In the next section \ref{sec:RFIM}, 
we introduce the three states RFIM and 
explain the thermodynamic properties including the critical phenomena such as phase transitions. 
In section \ref{sec:prediction}, 
we construct a prediction formula 
based on the model introduced in the previous section \ref{sec:RFIM}. 
We introduce `turnover'  as an order parameter 
to characterize the crisis. 
In section \ref{sec:sim}, 
we carry out computer simulations with the assistance of 
empirical data set to check the usefulness of our approach. 
The last section \ref{sec:summary} is concluding remark. 
 \section{There states random field Ising model}
 \label{sec:RFIM}
 In this paper, we extend the prediction model based on Ising model 
 given by \cite{Kaizoji2000,IHSI} by means of 
 three states random field Ising model. 
 Before we construct the prediction model for financial time-series, 
 we consider the thermodynamics of the following 
 Hamiltonian (energy function) that describes decision makings of $N$ traders 
 (each of the traders is specified by a label $i=1,\cdots,N$). 
\begin{equation}
\mathcal{H} (\bm{S}) = 
-\frac{J}{N}
\sum_{i,j=1}^{N}S_{i}S_{j} -h \sum_{i=1}^{N}\sigma (t) S_{i}-\mu \sum_{i=1}^{N}|S_{i}|
\label{eq:model}
\end{equation}
where each spin $S_{i}\,\, (i=1,\cdots,N)$ can take 
$\pm 1$ and $0$, 
and here we assume that all traders are located 
on a complete graph (they are fully connected). 
We should keep in mind that in the previous studies 
\cite{Kaizoji2000, IHSI}, 
a spin $S_{i}$ takes only $+1$ (buy) and 
$-1$  (sell). 
However, 
in our model system, 
besides $\pm 1$, 
$S_{i}$ can take $0$ which means that 
the trader $i$ takes a wait-and-see attitude (stays). 
Namely, 
\begin{equation}
S_{i} = 
\left\{
\begin{array}{cc}
+1 & (\mbox{buying}) \\
0 & (\mbox{staying}) \\
-1 & (\mbox{selling})
\end{array}
\right.
\label{eq:Ising}
\end{equation}
The first term in the right hand side of 
(\ref{eq:model}) causes the collective behavior of the traders 
because the Hamiltonian (\ref{eq:model}) decreases 
when all traders tend to take the same decision. 
In this sense, 
the first term is regarded as endogenous information for the traders. 
On the other hand, the second term 
denotes the exogenous information 
which is a kind of market information available for all traders. 
Here one can choose 
the following market trend during the past $\tau$-steps as $\sigma (t)$. 
\begin{equation}
\sigma (t) = 
\frac{q (t) -q (t-\tau)}{\tau}
\end{equation}
where $q (t)$ denotes a real price at time $t$. 
The third term appearing in the right hand side of 
equation (\ref{eq:model}) controls 
the number of traders who are staying at the moment $t$. 
From the view point of 
spin systems, 
$\sigma (t)$ is regarded as a `random field' on each spin 
because the $\sigma (t)$ might obey a stochastic process. 
Therefore,  the spin system described by (\ref{eq:model}) should be 
refereed to as {\it random field Ising model (RFIM)}. 
Obviously, 
the parameter $\mu$ is regarded as `chemical potential' in 
the literature of physics. 
For 
$\mu \gg 1$, 
most of the traders take `buying' or `selling' 
instead of `staying' 
from the view point of minimization of 
the Hamiltonian (\ref{eq:model}). 
In the limit of $\mu \to \infty$, 
the fraction of traders who take $S_{i}=0$ vanishes, 
namely, 
the system is identical to the conventional Ising model \cite{Kaizoji2000} in this limit. 
As we will see later, 
the set of parameters (what we call `hyper-parameters')  $(J,h,\mu)$ should be estimated (learned) from the past time-series. 

In this paper, we shall focus on 
the modification by means of the above three states RFIM.  
We investigate to what extent the 
prediction performance is improved. 
Moreover, 
we attempt to quantify the number 
of traders who are staying 
at the crushes in order to characterizing 
the financial crisis. 
\subsection{Equations of state}
To make a link between the prediction model and 
statistical physics of the three states RFIM, 
we should investigate the equilibrium state described by 
the Hamiltonian (\ref{eq:model}) at unit temperature. 
According to 
statistical mechanics, 
each microscopic state 
$\bm{S}=(S_{1},\cdots,S_{N})$ of the Hamiltonian (\ref{eq:model}) obeys 
the distribution 
${\exp}[-\mathcal{H}(\bm{S})]/Z$, where the normalization 
constant $Z=\sum_{\bm{S}=0,\pm 1}{\exp}[-\mathcal{H}(\bm{S})]$ is refereed to as {\it partition function} and it 
is given by 
\begin{equation}
Z = 
\sum_{\bm{S}=0,\pm 1}
{\exp}
\left[
\left(
\sqrt{\frac{J}{2N}}
\sum_{i=1}^{N}S_{i}
\right)^{2}
+h \sum_{i=1}^{N}\sigma (t) S_{i}
+
\mu \sum_{i=1}^{N}|S_{i}|
\right]
\end{equation}
where we defined 
$\sum_{\bm{S}=0,\pm 1} (\cdots) = 
\sum_{S_{1}=0,\pm 1} \cdots \sum_{S_{N}=0,\pm 1}(\cdots)$. 
Here we should keep in mind that arbitrary two traders are connected 
each other. 
By using a trivial equality concerning the Gaussian integral 
\begin{equation}
{\rm e}^{\alpha^{2}} = 
\int_{-\infty}^{\infty}
\frac{dx}{\sqrt{2\pi}}
{\exp}
\left(-\frac{x^{2}}{2} + \sqrt{2}\, \alpha x
\right), 
\end{equation}
the system is reduced to 
a single spin `$S$' problem in the limit of $N \to \infty$ as 
\begin{eqnarray}
Z  & = &   
\int_{-\infty}^{\infty} da 
\int_{-\infty}^{\infty} \frac{d\tilde{a}}{2\pi/N}
\int_{-\infty}^{\infty} \frac{dm}{\sqrt{2\pi/JN}}
{\rm e}^{-\frac{NJ}{2}m^{2}+N\mu a -N\tilde{a}a}  
\left\{
\sum_{S=0,\pm 1}
{\rm e}^{[Jm+h \sigma (t)]S+\tilde{a}|S|}
\right\}^{N}  \nonumber \\
\mbox{} & \simeq  & {\exp}\left[
N\Phi (m,a,\tilde{a})
\right]
\label{eq:Z}
\end{eqnarray}
where we used the saddle point method to evaluate the integrals 
with respect to $a,\tilde{a}$ and $m$. 
$\Phi (m,a,\tilde{a})$ appearing in the final form (\ref{eq:Z}) 
is regarded  as 
a free energy density and it is given by 
\begin{equation}
\Phi (m,a,\tilde{a})  =   
-\frac{J}{2}m^{2} +\mu a -a\tilde{a} + \log 
\left\{
1+{\rm e}^{\tilde{a}}
2\cosh [Jm + h \sigma (t)]
\right\}. 
\end{equation}
Then, the saddle point equation $\partial \Phi/\partial m=0$ 
leads to 
\begin{equation}
m = \frac{1}{N}\sum_{i=1}^{N}S_{i}= 
\frac{{\rm e}^{\tilde{a}}2\sinh [Jm+ h \sigma (t)]}
{1+{\rm e}^{\tilde{a}}2 \cosh [Jm + h \sigma (t)]}. 
\label{eq:m2}
\end{equation}
Apparently, 
the above $m$ stands for `magnetization' 
in the literature of statistical physics, 
however, 
as we will see later, 
it corresponds to 
the `return' in the context of time-series prediction for the price. 
This is because 
the number of buyers is larger than that of the sellers if the $m$ is positive, 
and as the result, the price increases definitely. 
Another saddle point equation 
$\partial \Phi/\partial \tilde{a}=0$ gives 
\begin{equation}
a = \frac{1}{N}
\sum_{i=1}^{N}|S_{i}| = 
\frac{{\rm e}^{\tilde{a}}2\cosh [Jm+ h \sigma (t)]}
{1+{\rm e}^{\tilde{a}}2 \cosh [Jm + h \sigma (t)]}. 
\label{eq:a2}
\end{equation}
It should be noticed that 
from $\partial \Phi/\partial a=0$, 
we have $\tilde{a}=\mu$. 
Hence, 
by substituting the $\tilde{a}=\mu$ into 
(\ref{eq:m2}) and (\ref{eq:a2}), 
we immediately have the following 
equations of state 
\begin{eqnarray}
m & = & 
\frac{2{\rm e}^{\mu}\,\sinh [Jm + h \sigma (t)]}
{1+2{\rm e}^{\mu}\,\cosh [Jm + h \sigma (t)]} 
\label{eq:m}\\
a & = & 
\frac{2{\rm e}^{\mu}\,\cosh [Jm + h \sigma (t)]}
{1+2{\rm e}^{\mu}\,\cosh [Jm + h \sigma (t)]}. 
\label{eq:a}
\end{eqnarray}
We should bear in mind that 
$a$ is a `slave variable' and it is completely determined 
by $m$. 
However, 
$a$ itself has an important meaning to characterize the market 
because 
the $a$ is regarded as 
the number of traders who are actually trading (instead of staying). 
In this sense, 
the $a$ could be `turnover' in the context of financial markets. 
In other words, 
the turnover $a$ is a measurement to quantify the activity of the market, 
and a large $a$ means high activity of the market. 
Strictly speaking, 
the $a$ could not be regarded as `turnover' because 
in our modeling, we assumed that 
each trader posts unit volume to the market. 
However, by introducing 
$v_{i}$ as volume for each trader $i$ and replacing the spin variables in (\ref{eq:model}) as 
$S_{i} \to v_{i}S_{i},\,\,\,v_{i} \in \mathbb{R}^{+}$, 
$a = (1/N)\sum_{i=1}^{N}|v_{i}S_{i}|$ 
is regarded as turnover in its original meaning. 

Obviously, from equations 
(\ref{eq:m}) and (\ref{eq:a}), 
the equation of state for the conventional Ising model \cite{Kaizoji2000,IHSI} 
is recovered in the limit of 
$\mu \to \infty$ as 
\begin{equation}
m = \tanh [Jm+h \sigma (t)], \,\,\, a=1. 
\end{equation}
In following, 
we analyze the above equations (\ref{eq:m})(\ref{eq:a}) 
to investigate the equilibrium properties of 
our model system. 
%
%
\subsection{Equilibrium states and phase transitions}
\begin{figure}[ht]
\begin{center}
\includegraphics[width=5.8cm]{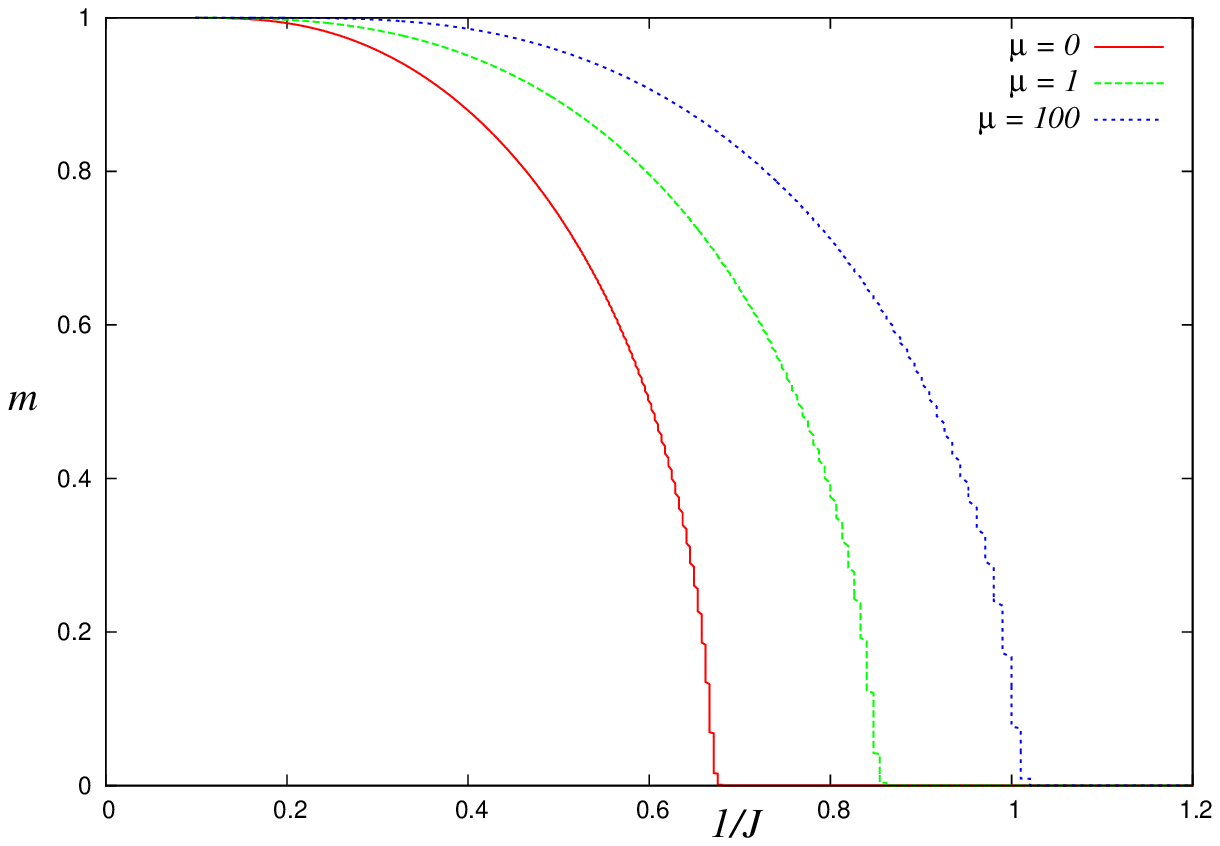}
\includegraphics[width=5.8cm]{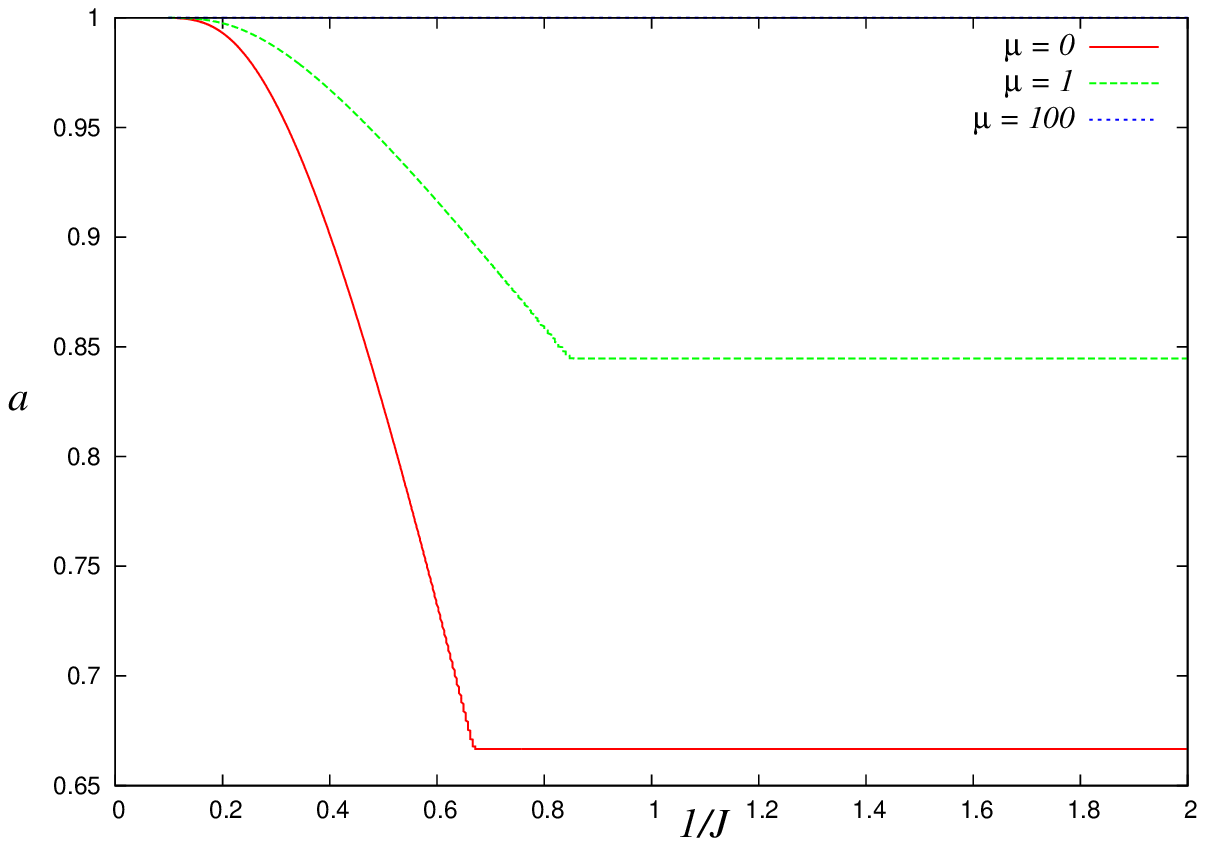}
\caption{\footnotesize 
$(1/J)$-dependence of 
magnetization $m$ and turnover $a$ for 
the case of $h=0$. 
At the critical point $(1/J)_{c} \equiv {2{\rm e}^{\mu}}/
{(1+2{\rm e}^{\mu})}$, 
the second order phase transition takes place. 
}
\label{fig:fg_h0}
\end{center}
\end{figure}

We first consider the case of $h=0$ in (\ref{eq:model}) or  (\ref{eq:m}) and (\ref{eq:a}). 
For this case, we can solve the equations of states 
(\ref{eq:m})(\ref{eq:a}) numerically. We show the 
$(1/J)$-dependence of magnetization $m$ 
in Fig. \ref{fig:fg_h0} (left). 
From this panel, 
we find that the magnetization 
$m$ monotonically decreases as $1/J$ increases for arbitrary finite $\mu$ 
and it drops to zero at the critical point $(1/J)_{c}$. 
The critical point is dependent on the value of $\mu$. 
In order to investigate the $\mu$-dependence of 
the critical point $(1/J)_{c}$, 
we expand the right hand side of (\ref{eq:m}) 
up to the first order of $m$. 
Then, we have  
\begin{equation}
(1/J)_{c} = 
\frac{2{\rm e}^{\mu}}
{1+2{\rm e}^{\mu}}. 
\end{equation}
It should be noted that 
$(1/J)_{c}=1$ for the conventional Ising model \cite{Kaizoji2000,IHSI}  
is recovered in the limit of 
$\mu \to \infty$. 
\mbox{} 

We next plot the $(1/J)$-dependence of 
the turnover $a$ in 
the right panel of Fig. \ref{fig:fg_h0}.  
From this panel, 
we are confirmed that 
above the critical point, 
the turnover $a$ takes a constant value: 
\begin{equation}
a = \frac{{\rm e}^{\mu}}{1+{\rm e}^{\mu}}
\end{equation}
Here we should notice again that 
$a=1$ is recovered 
in the limit of $\mu \to \infty$, 
which means that 
there is no trader who is staying at the moment. 
\begin{figure}[ht]
\begin{center}
\includegraphics[width=5.8cm]{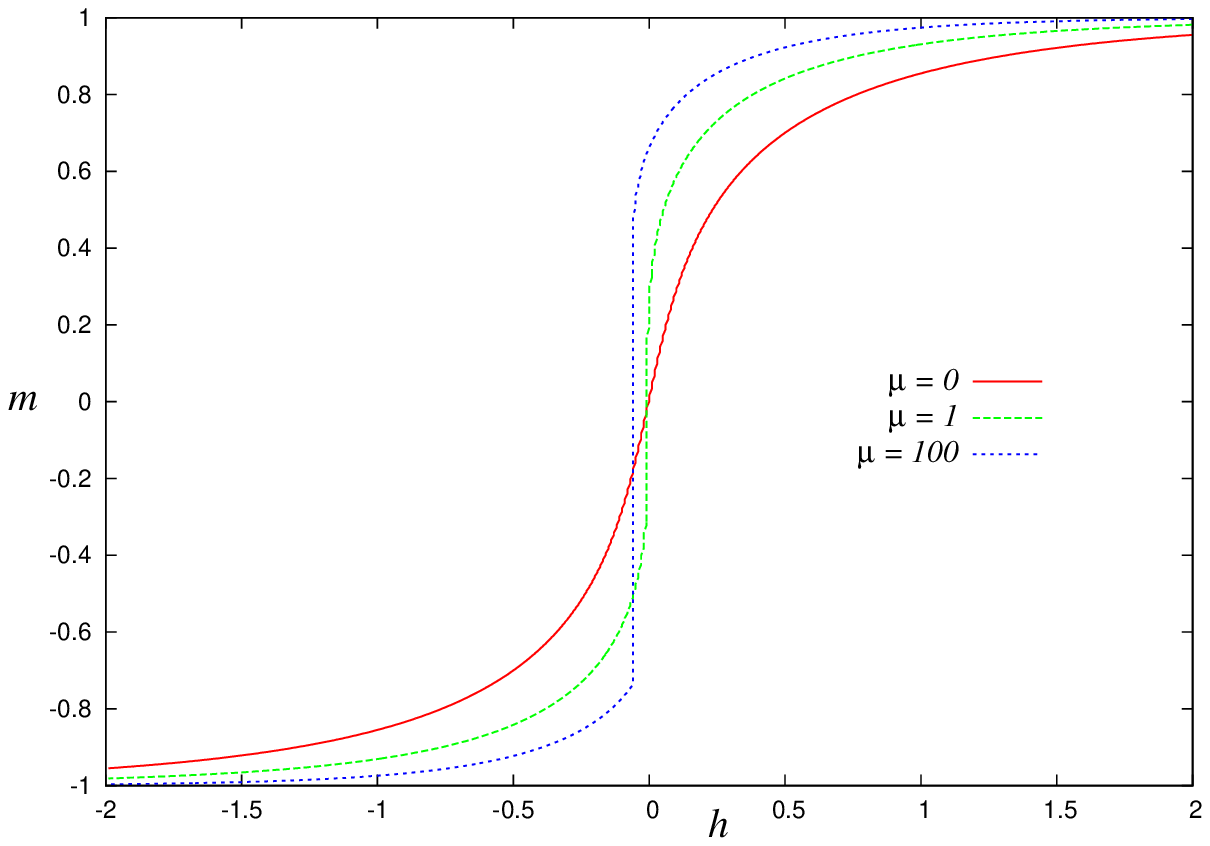}
\includegraphics[width=5.8cm]{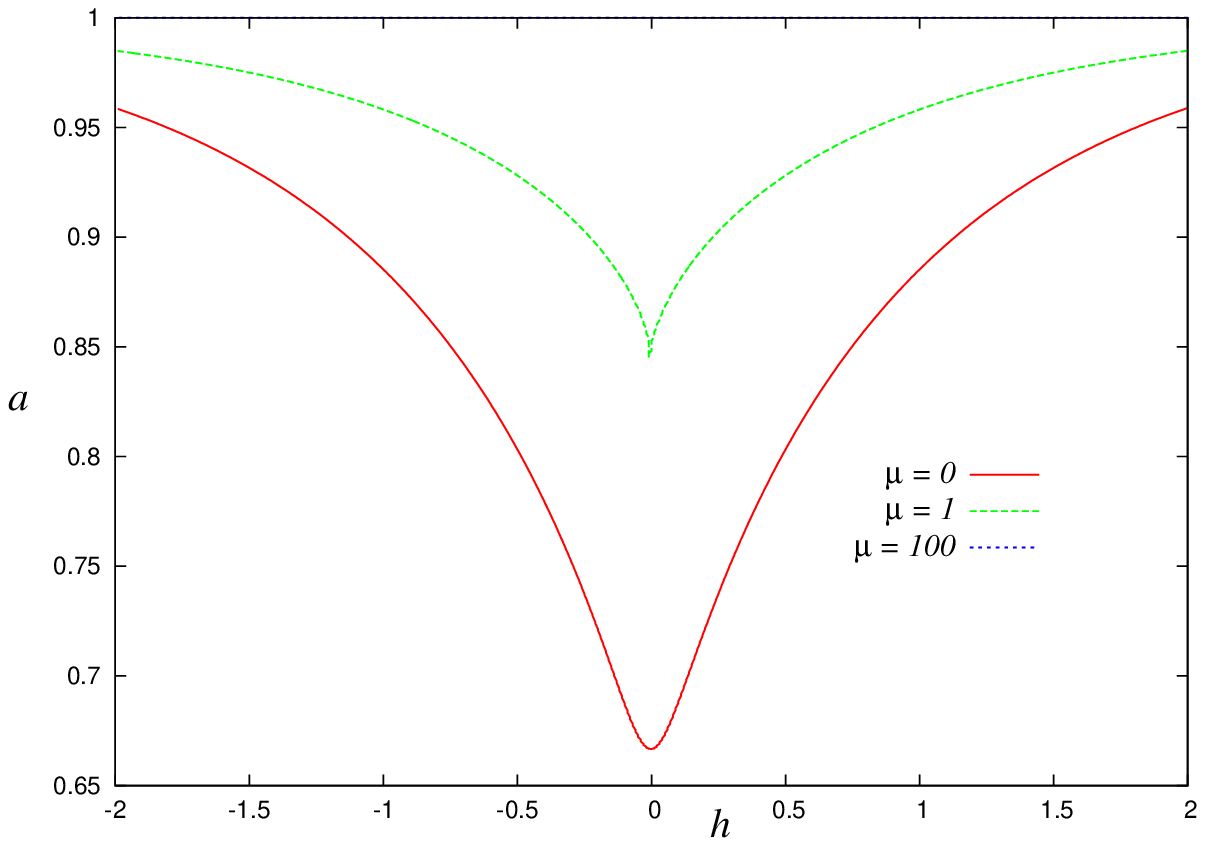}
\caption{\footnotesize 
Magnetization $a$ and 
turnover $a$ as a function of 
$h$. We plot each of them by keeping the $J$ as 
$J=1.2$. 
}
\label{fig:fg_J}
\end{center}
\end{figure}
\mbox{}

We next evaluate the behavior of 
magnetization $m$ as a function 
of $h$ keeping the value of $J$ as $J=1.2$ (we set $\sigma (t)=1$ for simplicity). 
Then, one observes from Fig. \ref{fig:fg_J} (left) that 
the system undergoes a first order phase transition 
which is specified by a transition between bi-stable states in the free energy, 
namely, the states $m>0$ and $m<0$. 
The critical values $m_{*}$ at the critical point is determined by 
\begin{equation}
1-m_{*}^{2} =
\frac{1}{4{\rm e}^{2\mu}-1}
\left\{
\frac{1-J(1-m_{*}^{2})}{2-2(1-m_{*}^{2})}
\right\} + \frac{1}{J}.
\end{equation}
We should notice that 
one recovers $m_{*}=\pm \sqrt{(J-1)/J}$, 
which is the result for the conventional Ising model \cite{Kaizoji2000}, 
in the limit of $\mu \to \infty$. 
For the case of 
$\mu < \infty$, 
the critical values $m_{*}$ is given by 
\begin{equation}
m_{*} = 
\left\{
\begin{array}{lc}
\pm \sqrt{\frac{J-1}{J}} & ( \mu  \geq \frac{1}{2} \log 
(
\frac{J^{2}+2}{8}
)) \\
\pm 
\frac{\sqrt{J^{2}-2(4{\rm e}^{2\mu}-1)}}{J} & ( \mu  < \frac{1}{2} \log 
(\frac{J^{2}+2}{8}
)
) 
\end{array}
\right.
\end{equation}
Then, the critical point $h_{c}$ is obtained as a solution of 
the following equation. 
\begin{equation}
\frac{1}{J}=
\frac{4{\rm e}^{2\mu} + 2{\rm e}^{\mu}\cosh(Jm_{*}+h_{c})}
{\{1+ 2{\rm e}^{\mu}\cosh(Jm_{*}+h_{c})\}^{2}}
\end{equation}
In the next section, taking into account the above 
equilibrium properties and 
phase transitions, 
we shall construct 
the prediction model 
based on the Hamiltonian (\ref{eq:model}) and 
evaluate the statistical performance by means of 
computer simulations with the assistance of empirical data analysis. 
\section{The prediction model}
\label{sec:prediction}
In this section, we construct our prediction model. 
Let us define $p (t)$ 
as the price at time $t$. 
Then, the return, which is 
defined as the difference between prices at successive two time steps 
$t$ and $t+1$, is given by 
\begin{equation}
p (t+1) -p (t) =\Delta (t). 
\label{eq:dyn}
\end{equation}
To construct the return $\Delta (t)$ from 
the microscopic view point, 
we assume that 
each trader ($i=1,\cdots,N$) buys or sells unit-volume, or stays at each time step $t$. 
Then, let us call the group of buyers as 
$\mathcal{A}_{+} (t)$, whereas 
the group of sellers is referred to as 
$\mathcal{A}_{-} (t)$. 
As we are dealing with three distinct states including 
`staying', we define the group of traders who are staying 
by $\mathcal{A}_{0} (t)$. 
Thus, the total volumes of buying, selling and staying are explicitly given by 
\begin{equation}
\phi_{+} (t) \equiv   \sum_{i \in \mathcal{A}_{+} (t)}1,\,\,
\phi_{-} (t) \equiv   \sum_{i \in \mathcal{A}_{-} (t)}1,\,\, 
\phi_{0} (t)  \equiv   \sum_{i \in \mathcal{A}_{0} (t)}1, 
\label{eq:psi}
\end{equation}
respectively. 
Apparently, 
the total number of traders should be conserved, namely, the condition 
$\mathcal{A}_{+} (t) + 
\mathcal{A}_{-} (t)  + \mathcal{A}_{0} (t) = N\, (\equiv \mbox{Total $\#$ of traders})$ holds. 

Then, the return $\Delta (t) $ 
is naturally defined by means of (\ref{eq:psi})  as 
\begin{equation}
\Delta (t)= \lambda (\phi_{+} (t)-\phi_{-} (t))
\end{equation}
where $\lambda$ is a positive constant. 
Namely, 
when the volume of 
buyers 
is greater than that of sellers, 
$\phi_{+} (t) > \phi_{-} (t)$, 
the return becomes positive $\Delta (t) >0$. 
As the result, the price should be increased at the next time step 
as $p(t+1)=p (t)+\Delta (t)$. 
\subsection{The Ising spin representation}
The making decision of each trader ($i=1,\cdots,N$) is now 
obtained simply by an Ising spin (\ref{eq:Ising}). 
The return is also simplified as  
\begin{equation}
\Delta (t)  =\lambda (\phi_{+} (t)-\phi_{-} (t)) = 
\lambda
\sum_{i=1}^{N}S_{i}^{(t)}  \equiv m_{t}
\end{equation}
where we set 
$\lambda=N^{-1}$ to 
make the return: 
\begin{equation}
m_{t} = 
\frac{1}{N}
\sum_{i=1}^{N}
S_{i}^{(t)}
\label{eq:update_pt}
\end{equation}
satisfying $|m_{t}|\leq 1$. 
Thus, $m_{t}$ corresponds to the so-called `magnetization' 
in statistical physics, and 
the update rule of the price is 
written in terms of the magnetization $m_{t}$ as 
\begin{equation}
p (t+1) = 
p (t) + m_{t}
\end{equation}
as we mentioned before. 
\subsection{The Boltzmann-Gibbs distribution}
It should be noticed that 
the state vectors of the traders: $\bm{S}=(S_{1},\cdots,S_{N})$ 
are determined so as to minimize 
the Hamiltonian (\ref{eq:model}) 
from the argument in the previous section. 
For most of the cases, 
the solution should be unique. 
However, 
in realistic financial markets, 
the decisions by traders should be much more `diverse'. 
Thus, here we consider 
statistical ensemble of traders $\bm{S}$ 
and define the distribution of 
the ensemble by $P(\bm{S})$. 
Then, we shall look for the 
suitable distribution 
which maximizes the so-called Shannon's entropy
\begin{equation}
H=-\sum_{\bm{S}=0,\pm 1}P(\bm{S}) \log P(\bm{S})
\end{equation}
under two distinct constraints: 
\begin{equation}
\sum_{\bm{S}=0,\pm 1}P(\bm{S})=1,\,\,
\sum_{\bm{S}=0,\pm 1}P(\bm{S})\mathcal{H}(\bm{S})=\mathcal{H}
\end{equation}
and we choose the $P(\bm{S})$ 
which minimizes the following functional $f \{P(\bm{S})\}$: 
\begin{eqnarray}
f\{P(\bm{S})\}  & = &    
-\sum_{\bm{S}=0,\pm 1}
P(\bm{S}) 
\log P(\bm{S})  - 
\lambda_{1}
\left(
\sum_{\bm{S}=0,\pm 1}
P(\bm{S}) -1
\right) \nonumber \\
\mbox{} & - & 
\lambda_{2}
\left(
\sum_{\bm{S}=0,\pm 1}
P(\bm{S}) 
\mathcal{H}(\bm{S})-\mathcal{H}
\right)
\end{eqnarray}
where $\lambda_{1}, 
\lambda_{2}$ are 
Lagrange's multipliers. 
After some easy algebra, we immediately obtain the solution 
\begin{equation}
P(\bm{S}) = 
\frac{{\exp}[-\beta \mathcal{H} (\bm{S})]}
{\sum_{\bm{S}=0,\pm 1} {\exp}[-\beta \mathcal{H}(\bm{S})]}
\label{eq:gibbs}
\end{equation}
where 
$\beta$ stands for the inverse-temperature. 
In following, we choose unit temperature $\beta=1$. 

Here we should assume that 
the magnetization as a return at time $t+1$ is given by 
the expectation of the quantity 
$(1/N)\sum_{i=1}^{N}S_{i}^{(t)}$ over the distribution 
(\ref{eq:gibbs}), that is, 
\begin{equation}
m_{t+1}= \sum_{\bm{S}^{(t)}=0,\pm 1}
\left\{
\frac{1}{N}
\sum_{i=1}^{N}S_{i}^{(t)}
\right\}
P(\bm{S}^{(t)}) = \frac{2{\rm e}^{\mu_{t}}\,\sinh [J_{t} m_{t} + h_{t} \sigma (t)]}
{1+2{\rm e}^{\mu_{t}}\,\cosh [J_{t}m_{t} + h_{t} \sigma (t)]} 
\end{equation}
where we defined 
\begin{equation}
\mathcal{H} (\bm{S}^{(t)}) = 
-\frac{J_{t}}{N}
\sum_{i,j=1}^{N}
S_{i}^{(t)}S_{j}^{(t)}
-h_{t}
\sum_{i=1}^{N}\sigma (t) S_{i}^{(t)} - 
\mu_{t}\sum_{i=1}^{N}|S_{i}^{(t)}|
\end{equation}
in (\ref{eq:gibbs}) 
and used (\ref{eq:m}) to evaluate the expectation.

Thus, we have the following prediction formula 
\begin{eqnarray}
p (t+1) & = & 
p (t) + m_{t} \\
m_{t} & = & 
\frac{2{\rm e}^{\mu_{t}}\,\sinh [J_{t}m_{t-1} + h_{t} \sigma (t)]}
{1+2{\rm e}^{\mu_{t}}\,\cosh [J_{t}m_{t-1} + h_{t} \sigma (t)]}  \\
J_{t+1} & = & 
J_{t} -
\eta \frac{\partial \mathcal{E}(J_{t},h_{t},\mu_{t})}{\partial J_{t}} \\
h_{t+1} & = & 
h_{t} -
\eta \frac{\partial \mathcal{E}(J_{t},h_{t},\mu_{t})}{\partial h_{t}} \\
\mu_{t+1} & = & 
\mu_{t} -
\eta \frac{\partial \mathcal{E}(J_{t},h_{t},\mu_{t})}{\partial \mu_{t}}
\end{eqnarray}
where we introduced the cost function $\mathcal{E}$ 
to determine the parameters 
$(J_{t},h_{t},\mu_{t})$ by means of 
gradient descent learning as 
\begin{equation}
\mathcal{E}(J_{t},h_{t},\mu_{t}) = 
\frac{1}{2}
\sum_{l=1}^{t}
\left[
\overline{
\Delta q (l)}
-
\frac{2{\rm e}^{\mu_{t}}\,\sinh [J_{t}  
\overline{
\Delta q (l-1)}
 + h_{t} \sigma (t)]}
{1+2{\rm e}^{\mu_{t}}\,\cosh [J_{t}
\overline{
\Delta q (l-1)}+ h_{t} \sigma(t)]} 
\right]^{2}
\end{equation}
and $\eta$ is a learning rate. 
To obtain the explicit form of the learning equations, 
we take the derivatives as 
\begin{eqnarray}
\frac{\partial \mathcal{E}}{\partial J_{t}} & = & 
-\sum_{l=1}^{t}
\left[
\overline{
\Delta q (l)}
-
\frac{2{\rm e}^{\mu_{t}}\,\sinh [J_{t}  
\overline{
\Delta q (l-1)}
 + h_{t} \sigma (t)]}
{1+2{\rm e}^{\mu_{t}}\,\cosh [J_{t}
\overline{
\Delta q (l-1)}+ h_{t} \sigma (t)]} 
\right] \nonumber \\
\mbox{} & \times & 
\frac{2{\rm e}^{\mu_{t}}
\overline{\Delta q (l-1)} 
\cosh [J_{t}
\overline{
\Delta q (l-1)}+ h_{t} \sigma (t)]
+4{\rm e}^{2\mu_{t}} \overline{\Delta q (l-1)}
}
{
\{
1+2{\rm e}^{\mu_{t}}\,\cosh [J_{t}
\overline{
\Delta q (l-1)}+ h_{t} \sigma (t)]
\}^{2}} \nonumber \\
\mbox{}\\ 
\frac{\partial \mathcal{E}}{\partial h_{t}} & = & 
-\sum_{l=1}^{t}
\left[
\overline{
\Delta q (l)}
-
\frac{2{\rm e}^{\mu_{t}}\,\sinh [J_{t}  
\overline{
\Delta q (l-1)}
 + h_{t} \sigma (t)]}
{1+2{\rm e}^{\mu_{t}}\,\cosh [J_{t}
\overline{
\Delta q (l-1)}+ h_{t} \sigma (t)]} 
\right] \nonumber \\
\mbox{} & \times & 
\frac{2{\rm e}^{\mu_{t}}
\sigma (t)
\cosh [J_{t}
\overline{
\Delta q (l-1)}+ h_{t} \sigma (t)]
+4{\rm e}^{2\mu_{t}} \sigma (t)
}
{
\{
1+2{\rm e}^{\mu_{t}}\,\cosh [J_{t}
\overline{
\Delta q (l-1)}+ h_{t} \sigma (t)]
\}^{2}} \\
\frac{\partial \mathcal{E}}{\partial \sigma_{t}} & = & 
-\sum_{l=1}^{t}
\left[
\overline{
\Delta q (l)}
-
\frac{2{\rm e}^{\mu_{t}}\,\sinh [J_{t}  
\overline{
\Delta q (l-1)}
 + h_{t} \sigma (t)]}
{1+2{\rm e}^{\mu_{t}}\,\cosh [J_{t}
\overline{
\Delta q (l-1)}+ h_{t} \sigma (t)]} 
\right] \nonumber \\
\mbox{} & \times & 
\frac{2{\rm e}^{\mu_{t}}
\sinh [J_{t}
\overline{
\Delta q (l-1)}+ h_{t} \sigma (t)]
}
{
\{
1+2{\rm e}^{\mu_{t}}\,\cosh [J_{t}
\overline{
\Delta q (l-1)}+ h_{t} \sigma (t)]
\}^{2}}. 
\end{eqnarray}
In the above expressions, 
$\overline{\Delta q (t)}$ is evaluated for 
the true price $q (t)$ by 
\begin{equation}
\overline{\Delta q (t)} \equiv  
\frac{1}{M}
\sum_{i=t-M+1}^{t}[q (i+1)-q (i)]. 
\end{equation}
By substituting 
$m_{t}$ and 
the set of 
parameters 
$(J_{t},h_{t},\mu_{t})$ into 
(\ref{eq:a}), 
we obtain the turnover $a$ at each time step as 
\begin{equation}
a_{t}= 
\frac{2{\rm e}^{\mu_{t}}\,\cosh [J_{t}m_{t} + h_{t} \sigma (t)]}
{1+2{\rm e}^{\mu_{t}}\,\cosh [J_{t}m_{t} + h_{t} \sigma (t)]}. 
\end{equation}
We should remember that 
we defined the exogenous information 
$\sigma (t)$ by the trend $\sigma (t) = 
[q (t)-q (t-\tau)]/\tau$ and here we choose 
$M=\tau$ to evaluate the trend and 
$\overline{\Delta q (t)}$. 
\section{Computer simulations}
\label{sec:sim}
\begin{figure}[ht]
\begin{center}
\includegraphics[width=11.5cm]{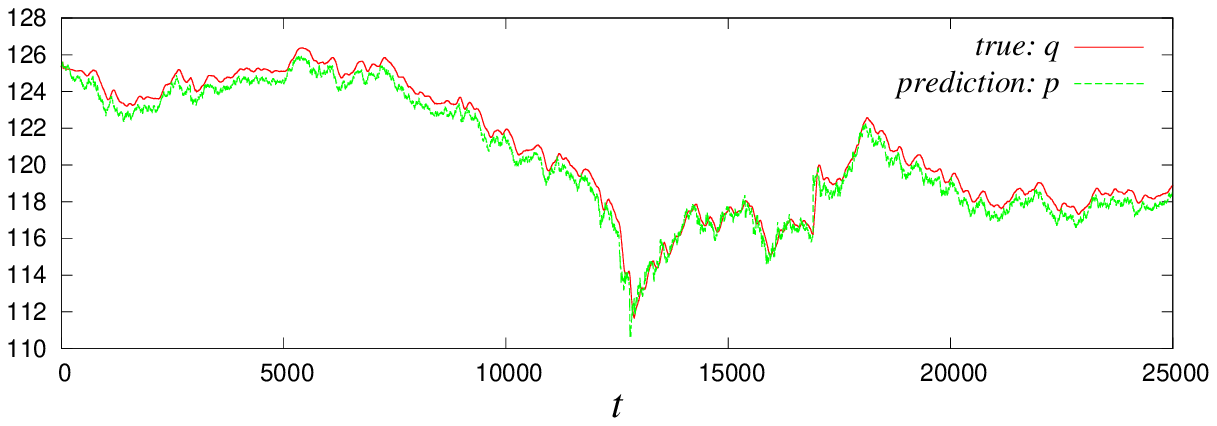}
\includegraphics[width=11.5cm]{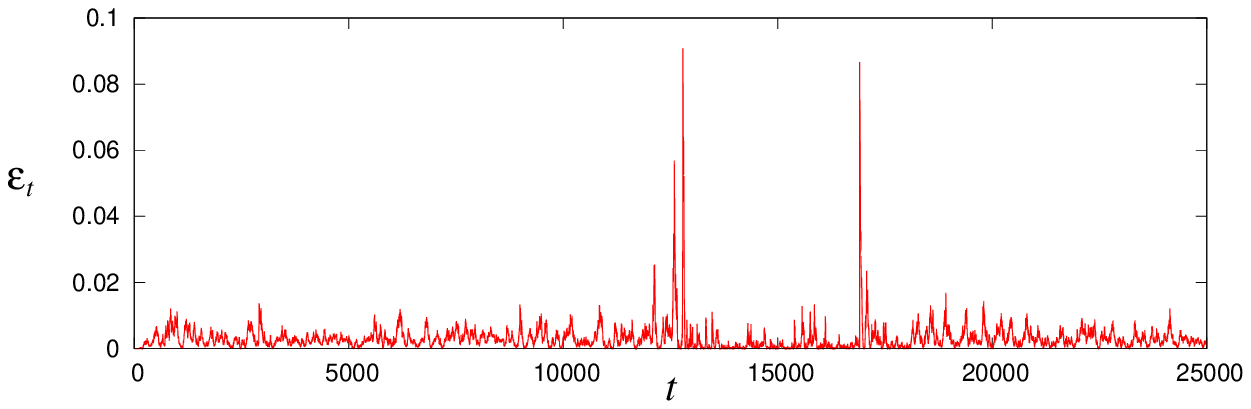}
\end{center}
\caption{\footnotesize 
The EUR/JPY exchange rate $q (t)$ (high frequency tick-by-tick data) from 
25th April 2010 to 13th May 2010, which was used in the reference \cite{IHSI} and 
the prediction $p (t)$. 
The lower panel shows 
the mean-square error 
$\varepsilon_{t} \equiv 
\{q (t)-p (t)\}^{2}/\max q (t)$. 
}
\label{fig:fg1}
\mbox{}\vspace{-0.4cm}
\end{figure}

In Fig. \ref{fig:fg1}, 
we show the true time-series $q (t)$ which contains a crush 
and the prediction $p (t)$ 
with mean-square error 
$\varepsilon_{t} \equiv \{q (t)-p (t)\}^{2}/\max q (t)$. 
The empirical true time-series $q (t)$ is chosen from 
EUR/JPY exchange rate (high frequency tick-by-tick data) from 
25th April 2010 to 13th May 2010 (it is the same data set as in the reference \cite{IHSI}). 
We set $\tau=M=100$ [ticks],  
and chose $J_{0}=0.1, h_{0}=0.6, \mu_{0}=0.1$ as the initial values of 
parameters. 
From these panels, 
we confirm that 
the mean-square error takes small value 
within at most several percent although 
the error increases around the crush. 
Thus, we might conclude that our three states RFIM 
works well on the prediction of financial data having a crush. 
\mbox{}

\begin{figure}[ht]
\begin{center}
\includegraphics[width=11.5cm]{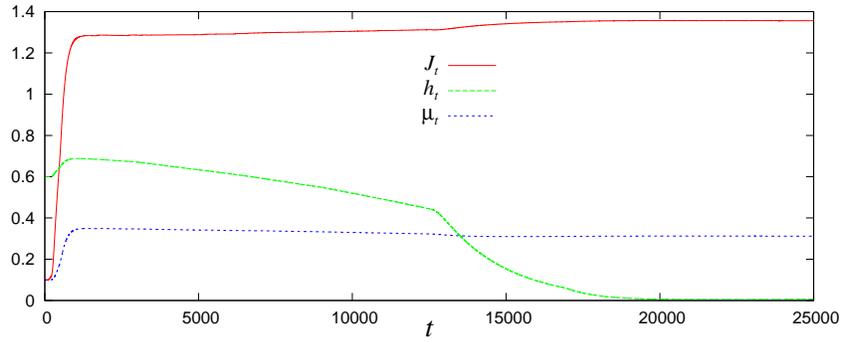}
\end{center}
\caption{\footnotesize 
Time-evolution of parameters $(J,h,\mu)$. 
The flow converges to the critical point. 
}
\label{fig:fg01}
\end{figure}

We next consider the flow of parameters 
$(J_{t},h_{t},\mu_{t})$ which evolve across the crush. 
The result is shown in Fig. \ref{fig:fg01}. 
From this figure, we clearly find that 
the strength of exogenous information $h_{t}$ drops to zero 
after the crush. Chemical potential $\mu_{t}$ and strength of 
endogenous information 
$J_{t}$ converge to 
$0.31$ and $1.35$, respectively. 
In our previous study \cite{IHSI}, 
as the critical point was $(1/J)_{c}=1$ for $\mu \to \infty$, 
the strength of endogenous information $J$ converged to the critical value $1$. 
However, 
in the three states RFIM 
with $h=0$ and $\mu=0.31 <\infty$, 
the critical point is sifted to $(1/J)_{c}=1/1.35 \simeq 0.74$
 (see Fig. \ref{fig:fg_h0} (left)). 
Therefore, in this simulation, 
these two parameters converge to 
the corresponding critical point $(J_{c},\mu_{c})=(1.35,0.31)$. 
From the result, we conclude that 
the system described by the Hamiltonian (\ref{eq:model}) 
automatically moves to the critical point after the crush.  
%

\begin{figure}[ht]
\begin{center}
\includegraphics[width=11.5cm]{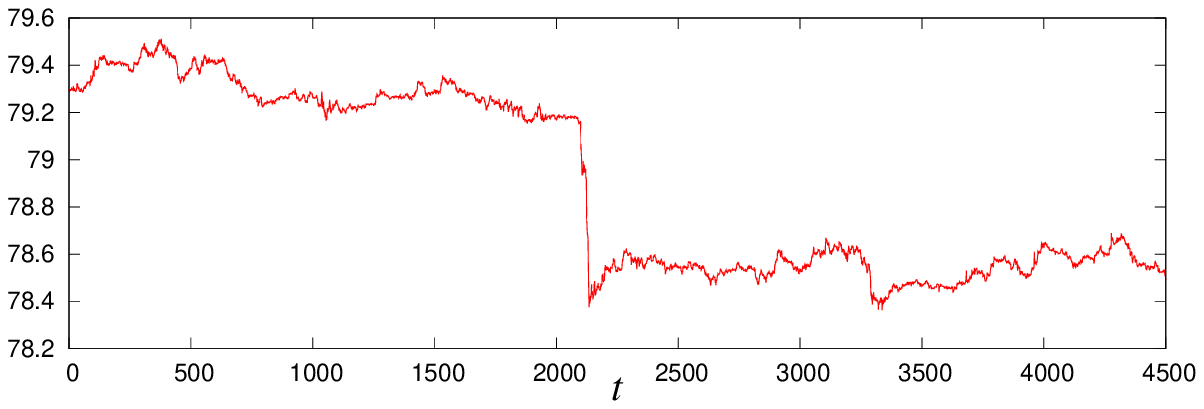}
\includegraphics[width=11.5cm]{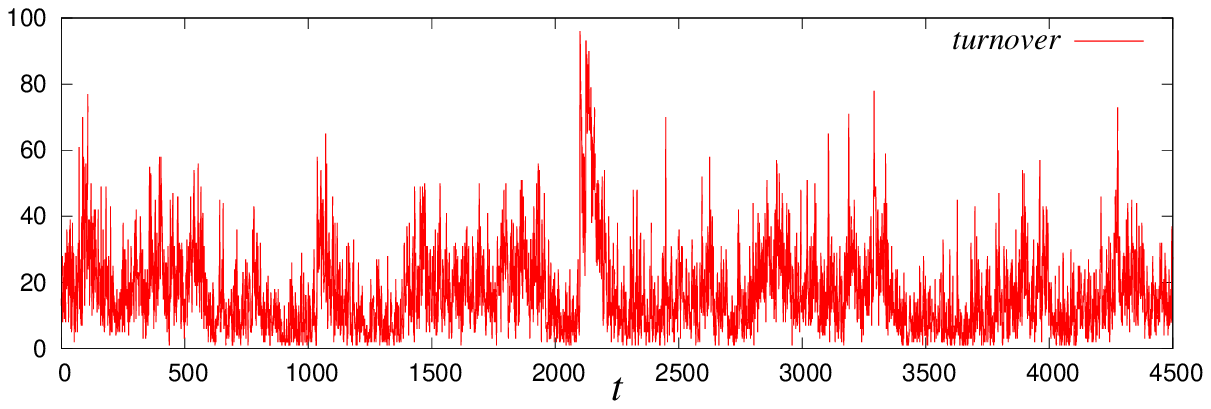}
\includegraphics[width=11.5cm]{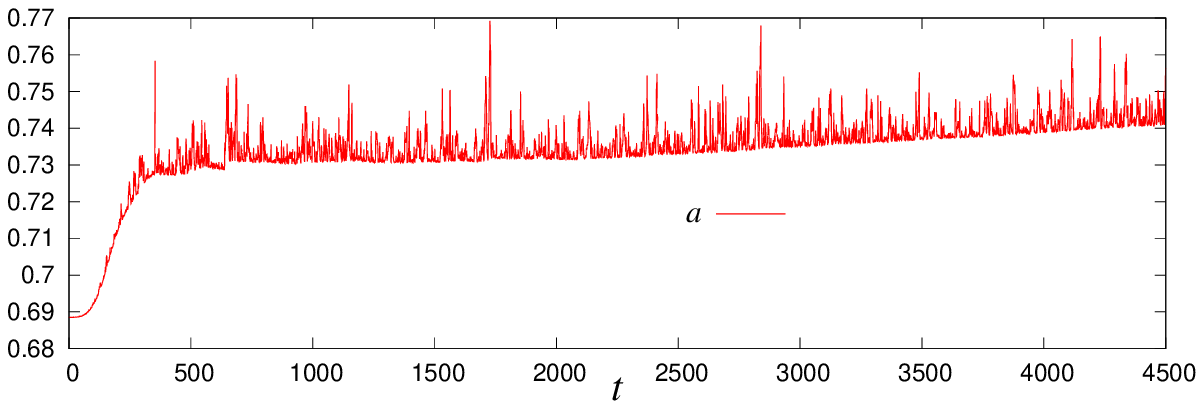}
\end{center}
\caption{\footnotesize 
USD/JPY exchange rate from 
12th August 2012 to 24th August 2012 as true time-series (the upper panel) and 
the true turnover (the middle panel). 
The lower panel shows 
the simulated turnover $a$ evaluated by our prediction model. 
}
\label{fig:fg2}
\end{figure}

We next utilize the USD/JPY exchange rate from 
12th August 2012 to 24th August 2012 as true time-series. 
It should be noted that the duration of 
this data is 1 minutes, hence, the data is not tick-by-tick data. 
In the simulation for this data set, 
we set $\tau=M=10$ [min]. 
We show the simulated turnover $a$ with the corresponding true value in Fig. \ref{fig:fg2}. 
From these panels, we find that 
the empirical data for the turnover increases instantaneously around the crush, 
whereas the simulated turnover $a$ does not show such striking feature 
although it possess a relatively large peak just before the crush. 
In order to convince ourselves that the simulated turnover $a$ can characterize 
the crush, we should carry out much more extensive simulations for various 
empirical data. It should be addressed as our future study. 
\subsection{Comparison with the conventional Ising model}
\begin{figure}[ht]
\begin{center}
\includegraphics[width=11.5cm]{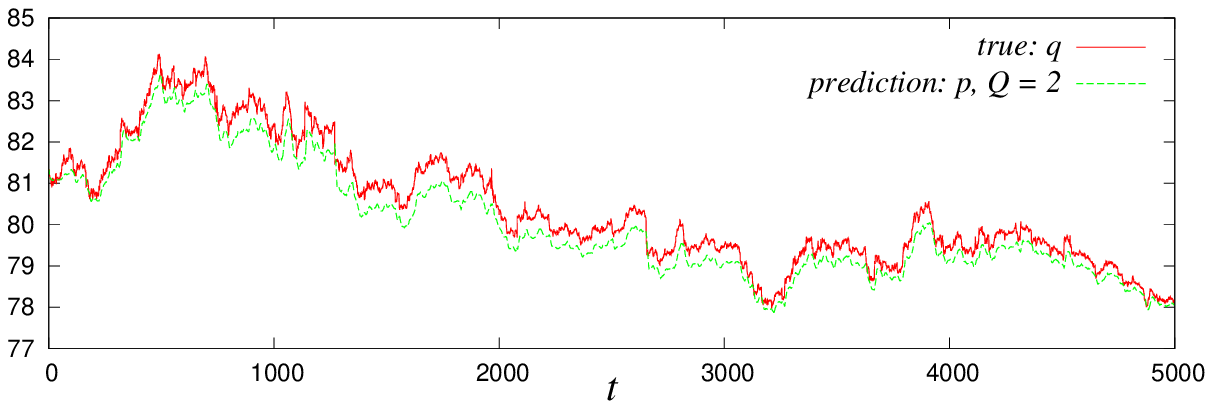}
\includegraphics[width=11.5cm]{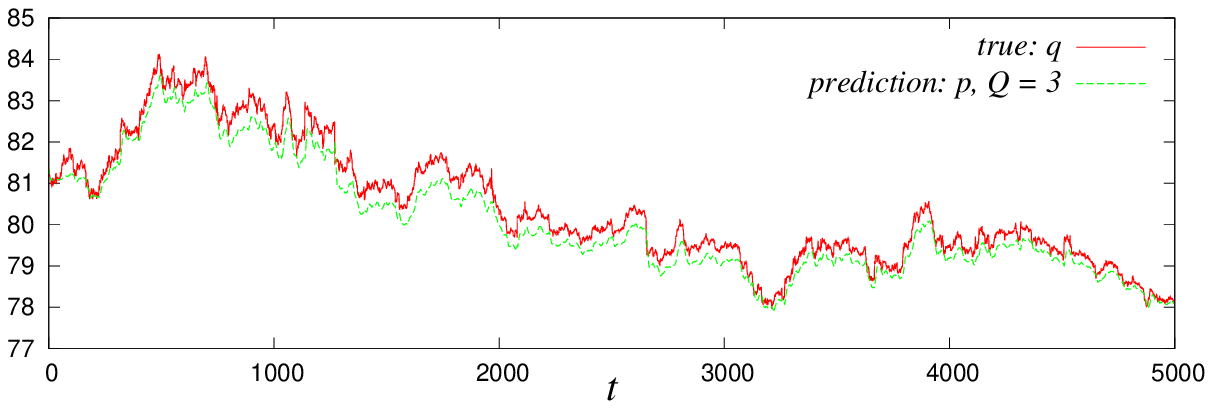}
\includegraphics[width=11.5cm]{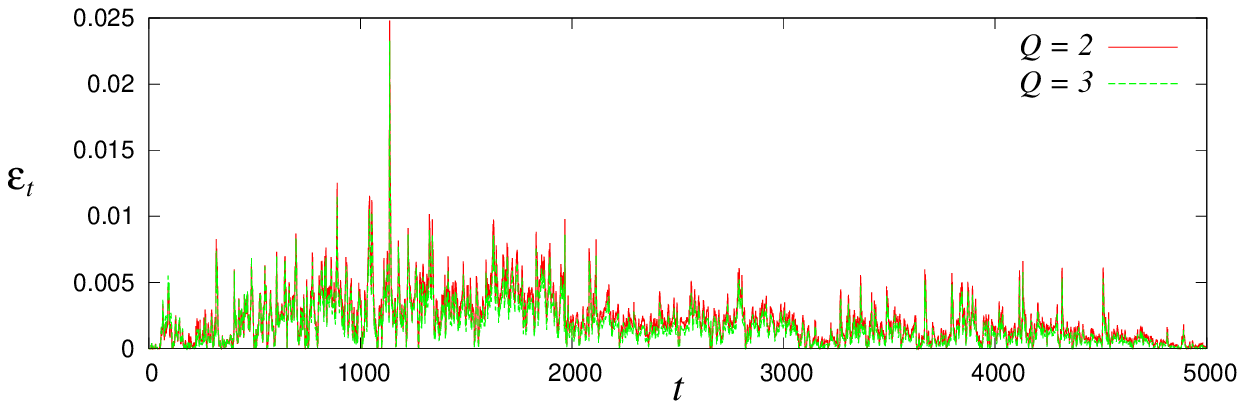}
\end{center}
\caption{\footnotesize 
Comparison with the conventional Ising model \cite{Kaizoji2000,IHSI}. 
The upper panel shows the result of the conventional Ising model, 
whereas in the middle panel, 
the result of three states RFIM is exhibited.  
The lower panel shows the corresponding mean-square errors. 
}
\label{fig:compare}
\end{figure}

Finally we compare our result with that of the conventional Ising model \cite{Kaizoji2000}. 
Here we used the USD/JPY exchange rate 
from 1st March 2012 to 31st July 2012, whose minimum duration 
is 30 minutes. 
We choose the width of the time-window $\tau=M=10$ [min]. 
The initial values of 
parameters are set to  $J_{0}=0.1, h_{0}=0.6$ for the conventional Ising model, whereas 
are chosen as $J_{0}=0.1, h_{0}=0.6, \mu_{0}=0.1$ for the three states RFIM. 
The results are shown in Fig. \ref{fig:compare}. 
From this figure as a limited result, we find that 
the performance of the prediction by the three states RFIM is 
superior to that of the conventional Ising model. 
\section{Concluding remark}
\label{sec:summary}
In this paper,  we extended the formulation of time-series prediction using Ising model given by 
Kaizoji (2001) \cite{Kaizoji2000} or Ibuki {\it et.\,al.}  (2012) \cite{IHSI} by means of 
three states RFIM. 
We found that 
the crisis could be `partially" characterized 
by the simulated turnover. 
We also confirmed that 
the three states $S_{i}=0,\pm 1$ in each trader's decision making apparently 
improves the statistical performance in the prediction. 
\subsection*{Acknowledgements}
This work was financially supported by 
Grant-in-Aid for Scientific Research (C) 
of Japan Society for 
the Promotion of Science, No. 22500195. 
The authors acknowledge 
Takero Ibuki, Shunsuke Higano and Sei Suzuki for fruitful discussion and useful comments. 
We thank organizers of {\it Econophysics-Kolkata VII}, especially, Frederic Abergel, Anirban Chakraborti, 
Asim K. Ghosh, Bikas K. Chakrabarti and Hideaki Aoyama. 



\begin{thebibliography}{99}


\bibitem{Reynolds}
C.W. Reynolds, 
{\it Flocks, Herds, and Schools: A Distributed Behavioral Model}, 
{\it Computer Graphics} {\bf 21}, 25 (1987). 


\bibitem{Makiguchi}
M. Makiguchi and J. Inoue, 
{\it Numerical Study on the Emergence of Anisotropy in Artificial Flocks: 
A BOIDS Modelling and Simulations of Empirical Findings},
{\it Proceedings of the Operational Research Society 
Simulation Workshop 2010 (SW10), CD-ROM}, pp. 96-102
 (the preprint version, arxiv:1004 3837) (2010). 


\bibitem{Kahneman}
D. Kahbeman and A. Tversky, 
{\it Econometrica} {\bf 47}, No.2, 263 (1979).  



\bibitem{ISI}
T. Ibuki, S. Suzuki and J. Inoue, 
{\it New Economic Windows (Proceedings of Econophysics-Kolkata VI)}, Springer-Verlag (Milan, Italy),  pp.239-259 (2013). 


\bibitem{IHSI}
T. Ibuki, S. Higano, S. Suzuki and J. Inoue, 
{\it Hierarchical information cascade: visualization and prediction of human collective behaviour at financial crisis by using stock-correlation}, 
{\it ASE Human Journal} {\bf 1}, Issue 2, pp. 74-87 (2012).



\bibitem{Mantegna}
R. Mantegna, 
{\it Euro. J. Phys. B} {\bf 11}, 193 (1999). 


\bibitem{Anirban}
J.-O. Onnela, A. Chakrabarti, K. Kaski, J. Kertesz and A. Kanto, 
{\it Phys. Rev. E}  {\bf 68}, 056110 (2003). 


\bibitem{MDS}
I. Borg and P. Groenen, 
{\it Modern Multidimensional Scaling: theory and applications}, 
Springer-Verlag, New York (2005). 



\bibitem{Yahoo}
http://finance.yahoo.co.jp/

\bibitem{Kaizoji2000}
T. Kaizoji, {\it Physica A} 
{\bf 287}, 493 (2000).


\end{thebibliography}
\end{document}